# Approximation for the Path Complexity of Binary Search Tree

**Abstract:** The complexity of an algorithm is an important parameter to determine its efficiency. They are of different types viz. *Time complexity*, *Space complexity*, etc. However, none of them consider the execution path as a complexity measure. *Ashok et al*, firstly proposed the notion of the *Path Complexity* of a program/algorithm, which defined based on the number of execution paths *as a* function of the input size. However, the notion of path complexity of the program, cannot apply to the *object-oriented* environment. Therefore, *Anupam et al*, has extended the notion of path complexity to the class as follows. The notion of the state of the class is defined based on structural representation (aka *state*) of the class. The class contains data members and data operations. It considers only those data operations that change the state of the class. The path complexity of the class is defined to be the number of valid input sequences, each of them containing valid data operations. *Anupam et al*, had applied this notion to the class *Stack*. However, the stack is basic and simple data structures. Therefore, in this research we have used a more complex class to understand the path complexity behavior in the object oriented environment. *Binary Search Tree* (BST) is one of the well known (and *more complex too*) data structure, which is useful in *sorting*, *searching*, *Traffic Engineering* and many more applications. We have analyzed the path complexity of the class BST based on the algorithms for insert and delete operations. Additionally, we have modified the *delete* operation to minimize the path complexity for the class BST.

**Keywords:** Algorithm, Approximation, Binary Search Tree, Complexity, Class complexity, Graph, Path complexity, Traffic Engineering, Software, Software Testing.

## 1. INTRODUCTION

A program is the lowest level unit in the software engineering environment. The program complexity measures used to give critical information about reliability and maintainability of a software system. Major software engineering effort spent during the maintenance and testing periods. Therefore, program complexity measures have been defined and analyzed in the literature [1-3] [7]. However, none of them consider the execution path as the complexity measures for the program. In [10], *Nejmeh et al.*, propose the notion of the path complexity measure. A survey of these approaches is given in *section 2.1*.

In today's object oriented environment every person, place, etc. can be represented as an object. The class contains data members and data operations for every object. One cannot extend the complexity measure of a program to find the complexity of the class. Therefore, the different complexity measure for the class, proposed in the literature [4-6] [9] [11-19]. A short survey of these approaches is given in *section 2.2*.

However, none of them consider the execution path as the complexity measure for the class. In [8], *Anupam* formally defined the notion of path complexity for the class and propose the complexity for the *Stack class* based on it. Since, stack is one of the basic data structures, which contains the *Push* and *Pop* operations; in this paper, we have studied the more complex data structure called *Binary Search Tree* (BST) and determined its path complexity. In addition, we have modified the *delete* algorithm in the BST class to reduce the path complexity. To the best of our knowledge, we are the first to analyze the path complexity of the BST class.

The rest of the paper is organized as follows. In section 2, we survey the existing approaches followed by our contribution. In section 3, we propose and analyze the path complexity measure for BST class. The conclusion and future scope of the proposed work is given in section 4. References are at the end. In Appendix A, we give an example for finding the complexity of BST.

## 2. RELATED WORK

### 2.1 Complexity Measures for the Program

*Lines of code* (LOC) is one of the basic and simple complexity measures. However, the LOC is independent of the logical complexity of a program [2]. A measure based on the Control flow graph (CFG) was suggested by Thomas McCabe [3] called *Cyclomatic complexity* measure. The McCabe's Cyclomatic complexity, measures the number of branches in a program. This measure gives same Cyclomatic complexity to the *if, while*, etc. In addition, it fails to consider the level of nesting for different control flow structures, e.g. same complexity for branches in *series* or *nested*. The advancement in Cyclomatic complexity is proposed in [21] [26].

A program complexity measure called the path complexity is based on the number of program execution paths. For a given input, program execution follows a sequence of program statements that is called the execution path. The path complexity $P(A, n)$ of program $A$ is defined as the total number of possible distinct execution paths over all inputs of size $n$ [7]. The path complexity of several *example programs* can be found in [22-25]. It is not a structural measure like McCabe's Cyclomatic complexity [3], because path complexity cannot be computed from the control flow graph only. Path complexity of a program is computed by analysis of the program. Path complexity is used to compare the complexities of two programs A1 and A2, which may be different even when both programs have isomorphic CFGs. In [10], *Nejmeh et al.,* proposed the NPATH measure to compute path complexity of the program. In [20], *Beth et al.,* evaluated the NPATH measure. However, it *fails* to identify the dependency between the two consecutive loops in which the output of the first loop affects the execution of the other loops.

### 2.2 Complexity Measures for the Class

A class contains data members and data operations. In [6], *John et al.* discussed four metrics as follows: *Weighted Methods per Class* (WMC), *Mean Method Complexity* (MMC), *Standard Deviation Method Complexity* (SDMC) and *Number of Trivial Methods* (NTM). The WMC, MMC and SDMC are based on the Cyclomatic complexity measure. In WMC, the class complexity is computed as the sum of Cyclomatic complexities of each data operation in the class. Therefore, a class containing 30 empty data operations is more complex than a class containing one data operations with 29 branches. In MMC, the mean of Cyclomatic complexities of all the methods of the class is taken into account. SDMC is the variation of WMC. NTM is obtained by counting the number of data operations in the class with complexity equal to one. In [4], *Smith et al.* stated various problems for class complexity viz. data flow analysis, control flow analysis, inheritance, polymorphism. They suggest the need of developing new techniques for state based testing of the class. Turner and Robson [5] gave a state based approach for testing classes. The *state of the class* is defined based on the structural representation of the class. In the class, there are data operations such as search and display, which do not change the structure of the class. We consider only those data operations that change the structure of the class i.e. *valid operation*. In [9] [11-14], the authors use the object matrices concept of comparing the class. In [15-16] [19], the authors use function points [17] for the various parameters, but not the execution paths. In [18], *Tan et al.,* estimate the *Lines Of Code* using the attributes, relationships and entities in the class. However, none of the previous approaches take into account the execution path, which is one of the most efficient measure for software.

### 2.3 Path complexity of the stack class [8]

The notion of path complexity of the class is "The path complexity P (A, n) of the class A is defined as the number of valid input sequences of length n>0, consisting of valid n data operations such that each of them changes the state of the class."

A class stack is one of the basic data structures. *Push* and *Pop* are valid data operations for the class stack. *Push* operation stores an item and *Pop* operation removes the most recently stored item from the stack. In a class stack, *state* represents the number of elements in the stack. The data operation *pop* not allowed on an empty stack. It input the sequence of size $n$ i.e. $(j_1, j_2, j_3, \ldots, j_n)$, $j_i \in \{Push, Pop\}$ and $1 \leq i \leq n$. The input sequence $(j_1, j_2, j_3, \ldots, j_n)$ for a stack is valid only if the sub sequence $(j_1, j_2, \ldots, j_k), 1 \leq k \leq n$ has the number of pop operations at most equal to the number of the push operations at any point. Let $P(stack, n)$ denote the path complexity of a stack class.

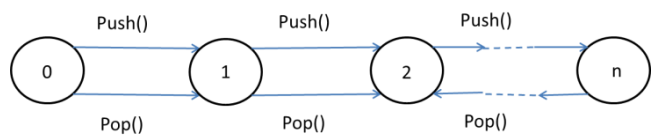

Fig. 1. State diagram of the class stack

Upper bound for the path complexity defined as follows. Assume $All\_Seq\ (n)$ as the set of all the possible input sequences of length $n$, $n \geq 1$. A class stack has two operations push and pop, and we have to select any one of them for each of the input operation then $P(stack, n) = 2^n$. However, pop operation is not allowed on empty stack, hence $P(stack, n) \leq 2^n$.

If out of $n$ valid operation, there are exactly $k$ pop operations then $0 \leq k \leq \lfloor\frac{n}{2}\rfloor$. Let $P(n, k)$ be the number of valid input sequences of length $n$ containing exactly k pop operations. In an input sequence of length $n$, the $n^{th}$ operation can be insert{push} or delete{pop}.

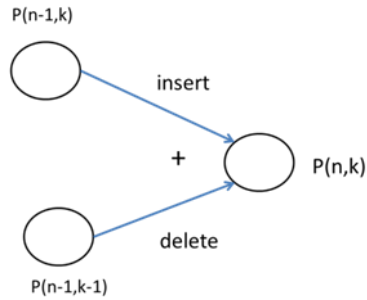

Fig. 2. Recurrence to obtain P(n,k)

As shown in Figure 2, we have the recurrence for P(n,k), given by $P(n, k) = P(n - 1, k - 1) + P(n - 1, k)$ i.e. an input sequence of size $n$ in $P(n, k)$ is obtained by addition of insert operation on P(n-1,k) or addition of delete operation on $P(n - 1, k - 1)$. If we make an insert or delete operation on a state $k$ then we have only one state, i.e. if insert than state $k + 1$ or if delete then state $k - 1$. If $k = 0$ and all the $n$ operations and there is one input sequence, hence $P(n, 0) = 1$. For $0 \leq k \leq \lfloor\frac{n}{2}\rfloor$

$$P(n, k) = \begin{cases} 1, & k = 0 \\ P(n - 1, k - 1) + P(n - 1, k), & k \neq 0 \end{cases}$$

The $P(n) = \sum_{k=0}^{\lfloor\frac{n}{2}\rfloor} P(n, k)$ denote the number of valid input sequences of size $n$ and $P(stack, n) = P(n)$.

The path complexity analysis of *Stack* is simple in the sense that an insert or delete operation results only in one state. For the data structure BST, an insert/delete results in more than one states.

### 2.4 Our Contribution

The approaches till now discussed the complexity of a program or class. To the best of our knowledge, none of the existing literature had taken the path complexity of Binary Search Tree (BST) into consideration. As the BST has various applications including sorting, searching etc. Therefore, it is necessary to obtain or determine the path complexity for a BST class. In this paper, we give the path complexity for the BST data structure. Thereafter we have proposed a new delete algorithm on the BST class to reduce the path complexity.

This measure consider the node as state and operation as transition, which relates to the graph algorithms. If we consider two different states as the source and destination then path complexity can lead to finding the shortest path from source to destination. Therefore, by looking at various applications of path complexity for BST we have proposed the approximation based on lower and upper bounds.

One can think of this measure as a possible futuristic solution to a *Traffic Engineering* (TE) model in which content provider has to select a route (i.e. path from one node to another node) to provide the service to the end users. From the available routes, the content provider select the path that minimizes the congestion and has low packet loss and high throughput with low latency. Therefore, using the path complexity measure, the content provider efficiently select the particular route among the other routes.

## 3. PATH COMPLEXITY OF A BST

A binary search tree (BST) is a binary tree in which each internal node $v$ has a distinct data value $e$ such that the values stored in the left subtree of node $v$ is less than $e$ and the values stored in the right subtree of $v$ is greater than $e$. Insert and delete operations are of our interest for the class BST. Here we assume that the class BST is initialized to an empty tree. Using the insert and delete operations, the path complexity of the class BST is analyzed. Later in the section, we have modified the algorithm for delete operation to facilitate path complexity analysis and given the bounds for path complexity of the class BST.

A state represents the structural representation of the BST. Structure is important in the sense that if we insert then we have to search position where we insert a new node and search depends upon the depth of the BST.

The state diagram for a BST with n=3 nodes is shown in Figure 3. The single arrow shows the insert operation and dashed arrow shows delete operation. The root node is colored black in each state. The initial state is the empty state. The stage $n$ contains all possible distinct BST states of order n.

| Notations | Meaning |
|---|---|
| Root($T_n$) | The root node of BST $T_n$ |
| Key(V) | The Value at node V |
| V.leftchild() | The Left child of Node V |
| V.rightchild() | The Right child of Node |
| NotLeafNode(V) | Returns true if Node V is not a leaf node |

| LeafNode(V) | Returns true if Node V is a leaf node |
| Successor (V) | The leftmost internal node in the right sub tree of $T_V$ |
| Predecessor (V) | The rightmost internal node in the left sub tree of $T_V$ |
| Empty(V) | Denotes the true if Node V is an empty node |
| NotEmpty(V) | Denotes the true if node V contains some value. |

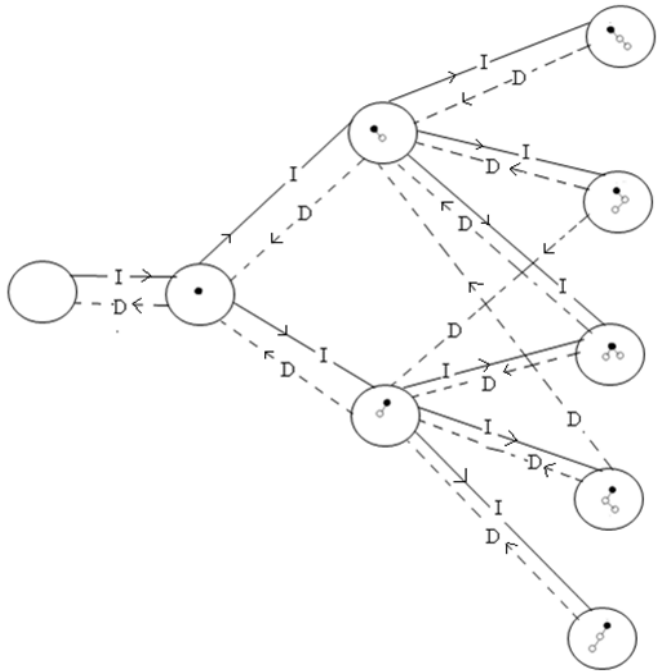

Fig. 3. Finite state diagram for the class BST for n=3 nodes

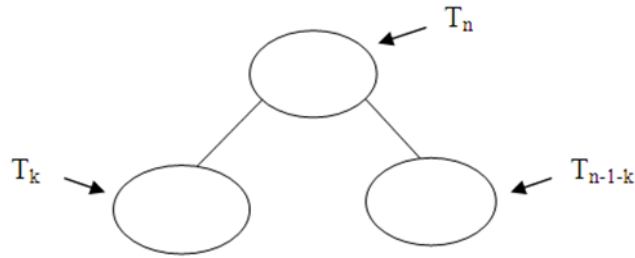

Fig. 4. BST of order *n*

***Proposition 2:*** Let $T_{n-1}$ be a BST of order $n-1$. An insert operation on $T_{n-1}$ results in one of $n$ possible BST $T_n$.

*Proof:* BST of order $n-1$ nodes has $n-1$ internal nodes and $n$ external nodes. Therefore, the new node after insert replaces any of $n$ external nodes. Therefore, there are $n$ distinct BST structures of order $n$.

Figure 5 shows a BST $T_4$ and five $T_5$'s obtained by inserting at leaf nodes. A label in external node indicate that by inserting node with label k, we got BST structure $a_k$ given at right side of arrow.

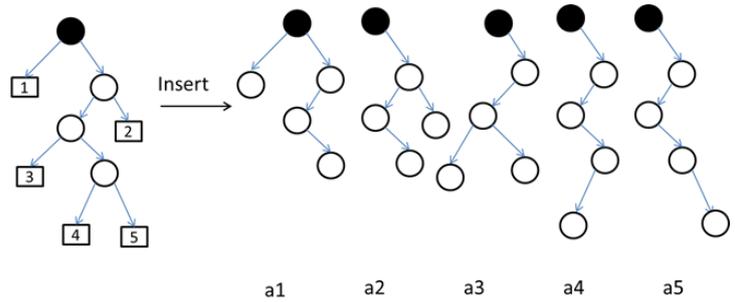

Fig. 5. Insert operation on BST

***Proposition 3:*** Let $T_n$ be a BST of order $n$ and let t be the number of leaf nodes of $T_n$, then a delete operation on $T_n$ yields at least $t$ distinct BSTs and at most $n-1$ distinct BSTs of order $n-1$.

*Proof:* If there are t leaf nodes then deleting a leaf node $v$ from $T_n$ yields a distinct BST $T_n - v$ and the conclusion follows. Then a delete operation on $T_n$ yields at least $t$ distinct BSTs where $t$ is the number of leaf nodes in $T_n$. To prove for at most $n-1$ distinct BSTs, we consider two cases based on whether deleting node, which had at most one child or two children and prove that in each case there were at least two nodes for which delete operation give isomorphic BSTs.

*Case 1:* If a node in BST $T_n$ has one child $v$ and $v$ is a leaf node, then there is a path from root to $v$. In this case deleting leaf node, which has no sibling give the same BST structure as deleting the parent of $v$.

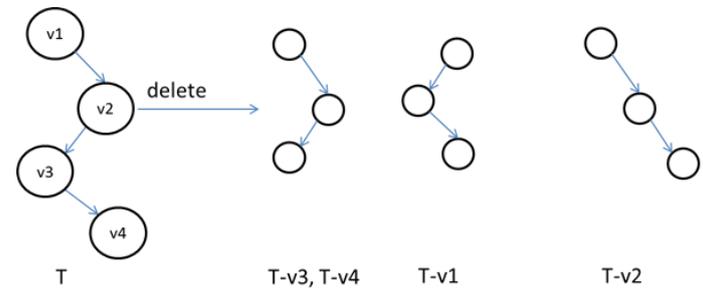

Fig. 6. Delete operation for case 1

*Case 2:* If a node *v* in BST $T_n$ has two children.

In this case deleting *v* is equivalent to deleting a successor of the node *v*.

**Table 1**: Number of distinct BST structures for n=1, 2 and 3 nodes

| n | $b_n$ | Possible BST structures | | | | |
|---|---|---|---|---|---|---|
| 0 | 1 | Empty Tree | | | | |
| 1 | 1 | I(1) | | | | |
| 2 | 2 | I(1),I(2) | I(2),I(1) | | | |
| 3 | 5 | I(1),I(2),I(3) | I(1),I(3),I(2) | I(2),I(1),I(3) I(2),I(3),I(1) | I(3),I(1),I(2) | I(3),I(2),I(1) |

Fig. 7. Delete operation for case 2

### 3.1 Path complexity analysis of the class BST

Let $S(n)$ be the number of valid input sequences for input length n with operations in {insert, delete}. If we assume that out of $n$ operations there are exactly $k$, $0 \leq k \leq \lfloor \frac{n}{2} \rfloor$ delete operations. Let $S(n,k)$ = number of input sequences of length n containing exactly k delete operations. $S(n) = \sum_{k=0}^{\lfloor \frac{n}{2} \rfloor} S(n,k)$, denote the number of valid input sequences of size $n$, where operations in {Insert, Delete}. $S(n)$ is identical to $P(stack, n)$ because in the class stack any input sequence had only one linear path in finite state diagram.

### 3.2 Determining path complexity of the class BST

It requires the following steps to be performed.

Generate all $S(n,k)$, valid input sequences of length $n$ containing $k$, $0 \leq k \leq \lfloor \frac{n}{2} \rfloor$ delete operations.

For each of sequence *s* generated, determine the total number of distinct execution paths $E(s)$ by using the Finite State Diagram. Here each input sequence is of length $n$ applied to the initial state.

Let $p(n,k)$ = Path complexity of all valid input sequences of length $n$ containing exactly $k$ delete operations, then $P(BST, n) = \sum_{k=0}^{\lfloor \frac{n}{2} \rfloor} p(n,k)$, where $p(n,k) = \sum_{s \in S(n,k)} E(s)$.

For $k = 0$ we have $n$ insert operations and an insert operation $i$, $0 \leq i \leq n$ has $i + 1$ paths to the next stage, therefore there are $n!$ distinct execution paths. After executing $n$ operations with $k$ delete operations, we are in stage $n - 2k$. If the n[th] operation is insert means we traverse from $n - 2k - 1$ stage to $n - 2k$ stage, so each state of stage $n - 2k - 1$ gives $n - 2k$ paths to stage $n - 2k$. The source of difficulty is, if n[th] operation is delete operation, so we had given multiplier $f(n,k)$ which is an expression in terms of $n$ and $k$, later in this chapter given bounds based on delete operation. The recurrence for $p(n,k)$ given by

$$p(n,k) = \begin{cases} n!, & k = 0 \\ p(n-1, k-1) * f(n,k) + p(n-1, k) * (n - 2k), & k \neq 0 \end{cases}$$

The path complexity of BST can be calculated from finite state diagram as the all possible valid input sequences of length $n$. Path complexity of the class BST for n=3 nodes is given in *Appendix A*.

**Observation 2:** Let $P1(BST, n)$ be the path complexity of the class BST, then $P_1(BST,n) \in O(n! * EXP(c_1 * n))$.

As shown in Table 2, *P1_approx(BST,n)* is the approximation of $P_1(BST,n)$ calculated using curve fitting tool of MATLAB®. $P1(BST,n) = 1.021 * EXP(0.135 * n) * n!$. So $P_1(BST,n) \in O(n! * EXP(c * n))$ where *c* is a constant value.

**Table 2.** Approximation of $P_1(BST,n)$ for *n*.

| n | $P_1(BST,n)$ | P1_approx(BST,n) |
|---|---|---|
| 1 | 1 | 1 |
| 2 | 3 | 3 |
| 3 | 9 | 9 |
| 4 | 43 | 42 |
| 5 | 239 | 241 |
| 6 | 1659 | 1,652 |
| 7 | 13231 | 13,239 |
| 8 | 121187 | 121,223 |
| 9 | 1243135 | 1,248,696 |
| 10 | 14,163,825 | 14,291,780 |

### 3.3 Modified DELETE Algorithm

To delete a node of BST there are two cases to be considered. i) Deleting a leaf node**:** To delete a node with no children is easy, we simply remove it from the tree. ii) Deleting a node with at least one child: Replace the node with its successor child or predecessor child (if successor child is empty), repeat this process till the deletion of the leaf node. Therefore, in this algorithm, we delete the leaf node in any case, and number of BST states is equal to the number of leaf nodes.

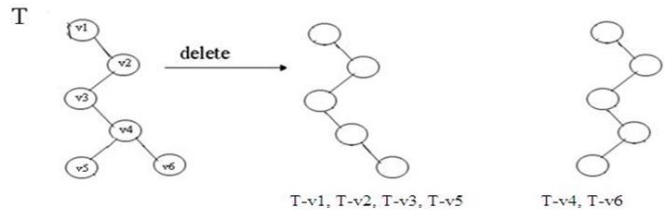

Fig. 8. Modified delete operation

| **Input** : A BST $T_n$ with *n* nodes and Node N that is to be deleted |
|---|
| **Output** : A BST $T_{n-1}$ with Node N deleted. |
| Delete2($T_n$, N) <br> {     // Check if tree is empty |

```
V = Root(T_n);
If (empty(V))
    Return NO SUCH KEY;
// Find the position of Node N
While (NotLeafNode(V) AND Key(N) != Key(V))
Do
    If (Key(N) > Key(V))
        V = V.rightchild();
    Else
        V = V.leftchild();
Done
If (LeafNode(V) AND Key(N) = Key(V))
    Delete Node V;
    Return T_{n-1};
If (Key(N) != Key(V))
    Return NO SUCH KEY;
While (NotLeafNode(V))
Do
    If (NotEmpty(V.rightchild()))
        Key(V)=Key(Successor(V));V=Successor (V);
    Else
        Key(V)=Key(Predecessor(V));V=Predecessor(V);
Done
Delete Node V; // V is a leaf node.
Return T_{n-1};
} // End Delete;
```

**Observation 3:** Let $P_2(BST,n)$ be the path complexity of the class BST, then $P_2(BST,n) \in O(n!*EXP(c_1*n))$ and $P_2(BST,n) \leq P_1(BST,n)$.

P2_approx(BST,n) is the approximation of $P_2(BST,n)$, calculated using curve fitting tool of MATLAB®. $P2(BST,n) = 1.109*EXP(0.09822*n) * n!$. So $P_2(BST,n) \in$

O(n!*EXP(c*n)) where *c* is constant value. Last column in Table 3 give the ratio of P₁(BST,n) to P₂(BST,n). From the ratio as it follows that P₂(BST,n) ≤ P₁(BST,n).

**Table 3.** Approximation of P2(BST,n) for *n*.

| n | P₂(BST,n) | P2_approx(BST,n) | P₁(BST,n) | $\frac{P_1(BST,n)}{P_2(BST,n)}$ |
|---|---|---|---|---|
| 1 | 1 | 1 | 1 | 1.00 |
| 2 | 3 | 3 | 3 | 1.00 |
| 3 | 9 | 9 | 9 | 1.00 |
| 4 | 41 | 39 | 43 | 1.05 |
| 5 | 219 | 217 | 239 | 1.09 |
| 6 | 1,447 | 1,439 | 1,659 | 1.15 |
| 7 | 11,081 | 11,116 | 13,231 | 1.19 |
| 8 | 97,533 | 98,108 | 121,187 | 1.24 |
| 9 | 965,631 | 974,097 | 1,243,135 | 1.29 |
| 10 | 10,634,115 | 10,746,292 | 14,163,825 | 1.33 |

If we make the insert operation from stage *n* than each of the state in stage *n* has exactly *n+1* execution paths mapped to the next stage, but when we have the delete operation than each of the state in stage *n* has *1* to *n-1* execution paths mapped to the next stage.

### 3.4 Bounds on the Path Complexity of the class BST

**Lower bound**: $P_{LB}(n) = \sum_{k=0}^{\lfloor \frac{n}{2} \rfloor} P_{LB}(n,k)$, where

$$P_{LB}(n,k) = \begin{cases} n!, & k = 0 \\ P_{LB}(n-1, k-1) * \frac{1+\lceil \frac{n-2k+1}{2} \rceil}{2} + P_{LB}(n-1, k) * (n-2k), & 0 < k \leq \lfloor \frac{n}{2} \rfloor \end{cases}$$

A BST $T_{n-2k+1}$ can have at most $\lceil \frac{n-2k+1}{2} \rceil$ leaf nodes. If $n^{th}$ operation is delete and let *t* be the number of leaf nodes, then from each of state in stage $n - 2k + 1$, there are at least *t* paths to stage $n - 2k$ ( by Proposition 3). So here we had taken average number of leaf nodes in BST $T_{n-2k+1}$ as a multiply factor to $P_{LB}(n-1, k-1)$.

**Upper bound**: $P_{UB}(n) = \sum_{k=0}^{\lfloor \frac{n}{2} \rfloor} P_{UB}(n,k)$, where

$$P_{UB}(n,k) = \begin{cases} n!, & k = 0 \\ (P_{UB}(n-1, k-1) + P_{UB}(n-1, k)) * (n-2k), & 0 < k \leq \lfloor \frac{n}{2} \rfloor \end{cases}$$

Delete operation take at most $n - 2k$ paths as we delete from stage $n - 2k + 1$(by Proposition 3).

## 4. CONCLUSION AND FUTURE WORK

This paper describes a measure of program complexity called path complexity that was proposed in [7]. Using the notion of path complexity, the path complexity of the class is defined in [8]. In this paper, the path complexity of the class BST is analyzed. We have given analysis based on the insert and delete algorithms of Binary Search Tree, and later we modify the delete algorithm to reduce the path complexity. The analysis of other data structures remain to be done for better understanding of the path complexity measure.

## 5. REFERENCES

[1] Donald E. Knuth, "Fundamentals algorithms 3rd edition, The art of computer programming", volume 1, pages 388-389, Addison-Wesley Professional, 1997.

[2] Rajib Mall, "Fundamentals of Software Engineering", Prentice Hall, Delhi, India 2001.

[3] Thomas J. McCabe, "A complexity Measure", IEEE transactions on software engineering, vol. Se-2, no. 4. Pages 308-320, Dec 1976.

[4] M.D. Smith and D. J. Robson, "Object-Oriented Programming - the Problems of Validation". In proc. IEEE Conference on software Maintenance, pages 272-281, 26-29 Nov. 1990.

[5] C.D.Turner and D.J. Robson, "The State-based Testing of Object-Oriented Programs". In proc. IEEE Conference on software Maintenance, pages 302-310, 27-30 Sep.1993.

[6] John Michura and Miriam A. M. Capretz, "Metrics Suite for Class Complexity", International Conference on Information Technology: Coding and Computing (ITCC'05), vol. 2, pages 404-409, 4-6 April 2005

[7] Ashok T. Amin and Naresh Jotwani, "On path complexity of Programs". DA-IICT, India, 2005.

[8] Anupam Mangal, "On Path Complexities of Heap sort algorithm and the class Stack". M.Tech. Thesis Report, DA-IICT, Gandhinagar, Gujarat,India,2007.

[9] Dr. Rakesh Kumar and Gurvinder Kaur. Article: Comparing Complexity in Accordance with Object Oriented Metrics. International Journal of Computer Applications 15(8):42–45, 2011. Published by Foundation of Computer Science.

[10] Brian A. Nejmeh, NPATH: a measure of execution path complexity and its applications, Communications of the ACM,v.31 n.2, p.188-200,Feb.1988.

[11] Chidamber S. R. and C. F. Kemerer, "A Metrics Suite for Object Oriented Design", IEEE Transactions on Software Engineering, Vol. 20, No. 6, June1994, pp. 476-493.

[12] Chucher N.I. and M.J. Shepperd, "Comments on a metrics Suite for Object-oriented Design" IEEE Transaction on Software Engineering, Vol. 21, No.3, 1995,pp. 263-265.

[13] Basili V. L., L. Briand and W. L. Melo, "A validation of object-oriented Metrics as Quality Indicators", IEEE Transaction Software Engineering. Vol.22, No. 10, 1996, pp. 751-761.

[14] Li W., "Another Metric Suite for Object-oriented Programming", The Journal of System and Software, Vol. 44, Issue 2, December 1998, pp. 155-162.

[15] Minkiewicz A., "The evolution of software size: A search for value," CROSSTALK, Vol. 22, No. 3,2009 pp. 23-26.

[16] Fraternali P., M. Tisi, and A. Bongio,"Automating function point analysis with model driven development," Proceedings of the Conference of the Center for Advanced Studies on Collaborative Research, Toronto, Canada, ACM Press, New York, 2006, pp. 1-12.

[17] Albrecht A. and J. Gaffney, "Software function, source lines of code and development effort prediction,"IEEE Transactions on Software Engineering, Vol. 9, 1983,pp.639-648.

[18] Tan H. B. K., Y. Zhao, and H. Zhang, "Estimating LOC for information systems from their conceptual data models," Proceedings of the 28th International Conference on Software Engineering, Shanghai, China, ACM Press, New York, 2006, pp. 321-330.

[19] Diev S., "Software estimation in the maintenance context," ACM Software Engineering Notes,Vol. 31, No. 2, 2006, pp. 1-8.

[20] R.Beth `McColl, James C. McKim Jr., Evaluating and extending npath as a software complexity measure, Journal of Systems and Software, Volume 17, Issue 3, March 1992, Pages 275-279, ISSN 0164-1212, 10.1016/0164-1212(92)90116-2.

[21] Lakshmanan, K.B.; Jayaprakash, S.; Sinha, P.K.; , "Properties of control-flow complexity measures," Software Engineering, IEEE Transactions on , vol.17, no.12, pp.1289-1295, Dec 1991. doi: 10.1109/32.106989

[22] Kamra M., "Analysis of Path Complexity", B. Tech. Project Report,DA-IICT, India, 2006

[23] Tushar Wadhwa., Ashok T. Amin." Analysis of Path Complexity of Programs", B. Tech. Project Report,DA-IICT, India, 2007

[24] Richa Gulati., Ashok T. Amin." Analysis of Path Complexity of Programs", B. Tech. Project Report,DA-IICT, India, 2007

[25] Shubham Saxena., Ashok T. Amin." Analysis of Path Complexity of Programs", B. Tech. Project Report, DA-IICT, India, 2008.

[26] Brian Henderson-Sellers, Yagna Raj Pant, June M. Vemer," Cyclomatic Complexity: theme and variations", Australasian Journal of Information Systems, Volume 1 Issue 1, pp:24-37, 1993

## APPENDIX A: Path complexity of the class BST for n=3 nodes

Execution Paths for input sequences (iii),(idi) and (iid) is shown in Figure 9. So the $P_1(BST,3) = p(3,0) + p(3,1) = 9$.

| n | k | Execution paths for input sequences (separated by comma) |
|---|---|---|
| 3 | 0 | E(iii)=6  so p(3,0)=6 |
| 3 | 1 | E(idi)=1, E(iid)=2 so p(3,1)=3 |

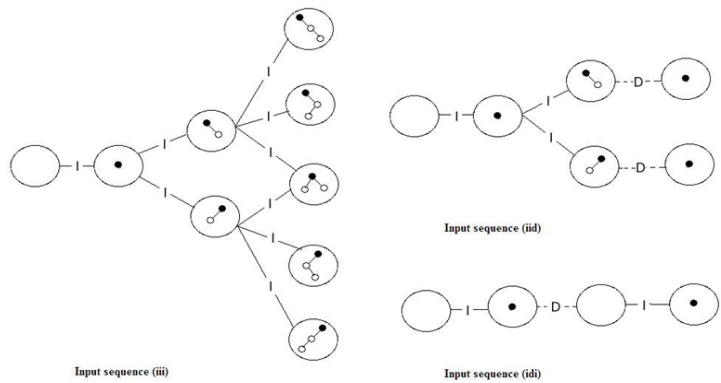

**Fig. 9.** Execution Paths for input sequences (iii),(iid) and (idi)